\documentclass[aip,chaos,subeqns.floatfix,onecolumn,amsmath,showpacs]{revtex4}
\usepackage{graphicx}
\usepackage{subfigure}
\usepackage{hyperref}
\begin{document}
\title{Extreme multistability: Attractor manipulation and robustness}
\author{ Chittaranjan Hens$^{1,2}$, Syamal K. Dana$^1$, Ulrike Feudel$^{3,4}$ }
\affiliation{$^1$CSIR-Indian Institute of Chemical Biology, Kolkata 700032, India}
\affiliation{$^2$Department of Mathematics, Bar-Ilan University, Ramat Gan, Israel}
\affiliation{$^3$Institute for Chemistry and Biology of the Marine Environment,
University of Oldenburg, Oldenburg, Germany}
\affiliation{$^4$ Institute for Physical Science and Technology ,University of Maryland, College Park, MD 20742-2431}
\date{\today}
\begin{abstract}
 The coexistence of infinitely many attractors is called extreme multistability in dynamical systems. In coupled systems, this phenomenon is closely related to partial synchrony and characterized by the emergence of a conserved quantity. We propose a general design of coupling  that leads to partial synchronization, which may be a partial complete synchronization or partial antisynchronization and even a mixed state of complete synchronization and antisynchronization in two coupled systems and, thereby reveal the emergence of extreme multistability. The proposed design of coupling has wider options and allows amplification or attenuation of the  amplitude of the attractors whenever it is necessary. We demonstrate that this phenomenon is robust to parameter mismatch of the coupled oscillators.
\end{abstract}
\maketitle
\begin{quotation}
{\bf  Multistability-the coexistence of many attractors  is an intrinsic property of many dynamical systems
in physics, chemistry, biology, and in ecosystems and  synthetic genetic networks. It has also been demonstrated in laboratory experiments,
such as lasers and electronics. Multistability reveals a rich diversity of stable states of a dynamical system and, as a consequence, such systems offer a great flexibility since they allow switching   from one stable state to another either by the influence of noise or other intrinsic control mechanism. This is, particularly, important for the survival of species in ecosystem. Hence, understanding  multistability and its control is an important issue. On the other hand, generating multistability is a challenge. 
We consider systems possessing not only a finite number of coexisting attractors but infinitely many of them. Such systems can be constructed by connecting two dynamical systems with an appropriate design of coupling even if the isolated systems do not show multistability. We  showed earlier that a partial complete synchronization is an essential condition for extreme multistability when a conserved quantity emerges.  Besides complete synchronization, in physical, chemical and biological systems, other forms of synchrony such as antisynchonization or mixed synchronization may exist. Here we establish that  extreme multistability can be realized with partial antisynchronization, and even under the condition of   mixed synchronization in two coupled systems, if we appropriately generalize the  strategy of the coupling design.  Additionally, we demonstrate that we can, in fact, increase and decrease the  amplitude of all attractors in a desired way,  whenever it is necessary. Furthermore, we establish the robustness of this phenomenon with respect to parameter mismatch of the coupled systems.
}
\end {quotation}

\section{Introduction}
\par The  coexistence of a multitude of different stable states is observed in many systems (cf. \cite{Feudel} for a review):
bistability in neurocortical  oscillations \cite{Foss,Freyer}, several possible overturning states  in the thermohaline ocean
circulation (THC) \cite{Rahmstorf}, multiple equilibria in a global ocean circulation \cite{Power}, different community compositions of species in ecosystems \cite{Huisman} and various gene expression in synthetic genetic networks \cite{Ullner}.
Multistablity is also seen in experiments  in lasers and  opto-electronic devices \cite{Masoller, Pisarchik}, in condensed matter physics \cite{Schwarz} and electronic circuits \cite{Lynch}.
From the theoretical point of view one can distinguish several system classes which possess coexisting attractors: weakly dissipative systems \cite{Feudel-et-al-96}, coupled systems \cite{Kaneko-93} or  time-delayed feedback systems \cite{Kim-97}.
Particularly in coupled systems the coexistence of multiple attractors is closely related to the coexistence of synchronized and desynchronized states. Multistability in coupled oscillators can be obtained (i) in cases where the single isolated oscillator already shows  multistability as in two coupled rotors \cite{Feudel-et-al} as well as (ii) in the case when the single oscillator possesses only one attractor as in coupled phase oscillators \cite{Astakhov}.
On the other hand, creating a large number of coexisting  attractors is an interesting issue to investigate and its control is an even more challenging task.
 It has been shown that one can get an arbitrarily large number of coexisting attractors in weakly dissipative systems by choosing a very small dissipation \cite{Feudel-et-al-96}. Another possibility is to consider a drive response system where the driver is a conservative system, i.e. for each initial condition of the driver, the driving force on the response system is different leading to even an infinity of attractors \cite{Lai-Grebogi}.
Alternatively, infinitely many attractors can be achieved when nonlinear systems are coupled in a specific way to show partial synchrony \cite{Sun, Ngonghala, Hens,Poria}.
 \par Theoretically and analytically, multistability has been investigated  in chemical reactions like the oscillatory Belousov-Zhabotinskii reaction \cite{Crowley}. Even more complex multistable behaviours  appears  in chlorite-thiosulfate reaction \cite{Orban-Epstein}. Despite  special care taken to ensure the same experimental conditions, this reaction exhibits different outcomes. One possible explanation of the experimental observation of seemingly infinitely many attractors has been offered by Wang et al. \cite{Wang} using  the  autocatalator model, where the inclusion of a buffer state creates the coexistence of different states. Later, Sun {\it et al} \cite{Sun} reported the existence of an infinite number of coexisting attractors, called   extreme multistability (EM) in  two coupled Lorenz systems by using a particular type of coupling. How to design a coupling scheme leading to extreme multistability in coupled oscillators, in a straightforward way, has been demonstrated in \cite{Hens} and later in \cite{Poria}. The emergence of extreme multistability has also been  evidenced  recently in an electronic experiment with two coupled R\"ossler oscillators \cite{Patel}. 
The main characteristics of extreme multistability has been studied in detail in \cite{Ngonghala}. This study reveals that two
properties are essential for the appearance of extreme multistability in two coupled $n$-dimensional nonlinear systems: (i) the complete synchronization (CS) of $n-1$ components of the two systems and (ii) {\it the emergence} of a conserved quantity $C$ in the long-term limit $t \to
\infty$. Note that in CS of two coupled systems, where all the $n$ components synchronize, the long term dynamics lies on a unique synchronization manifold \cite{Pecora, Kurths, Roy}. By contrast, in case of EM, only $n-1$ components synchronize completely, while the differences between the $n-th$ components of the two systems remain in a constant distance to each other. This constant distance $C$ defines a synchronization manifold in which the dynamics takes place. This conserved quantity $C$, which appears only in the long-term limit $t \to \infty$, depends on the initial conditions in a possibly complicated way. Since the value of $C$ can take any real number, the whole state space is foliated into infinitely many synchronization manifolds with at least one attractor existing in each manifold. 

\par  It is important to note, that the attractors in the synchronization manifold do not correspond to the usual definition of an attractor even in the Milnor sense \cite{Milnor}. According to Milnor, an attractor possesses a basin of attraction of positive Lebegue measure. As pointed out above, the whole state space is foliated into infinitely many such hypersurfaces as time goes to infinity. This can be considered as a foliation of the state space into infinitely many leafs. Within the each hypersurface or leaf we obtain a usual attractor as in any other dynamical system but only with respect to this leaf. In the direction perpendicular to the leaf we have marginal stability. That means that the invariant set in the leaf is in fact a {\it relative attractor} \cite{explain} since it is stable in all directions within the leaf or hypersurface but marginally stable in the direction perpendicular to it. Since the basin of attraction of each attractor consists in the simplest case of a $2n-1$ dimensional hypersurface, its Lebegue measure is zero in the $2n$ dimensional state space, but the attractor would have positive measure with respect to the hypersurface or leaf when considered as a relative attractor. To avoid confusion, we would like to stick with the notion of an "attractor" instead of a "relative attractor" since this phenomenon of extreme multistability has been discussed earlier in several papers (cf. \cite{Sun, Ngonghala}) and we would like to refer to those papers using their notation. Throughout the whole paper all "attractors" should be understood as "relative attractors". 

\par The general rule for defining the coupling function that successfully creates EM in two oscillators \cite{Hens} is based on
 Lyapunov function stability (LFS) \cite{Padmanaban}. Since different   types of synchronization like antisynchronization or mixed type of sychronization may appear in coupled
nonlinear systems, we  address here the question, if EM can exist for partial antisynchronization (AS) or in a state of mixed synchronization (MS) besides partial CS. In a state of AS, all the pairs of state  variables of the two oscillators are correlated by amplitude but are exactly opposite in phase ($\pi$) while in the MS state, some of the pairs of variables
are in CS and others in AS state. We make  an appropriate generalization of the definition of the coupling function to achieve this goal and show that the number of choices of the coupling is not restricted but widely open. \\
 An interesting benefit of our specific coupling for two oscillators is a possible amplification or attenuation \cite{Padmanaban} of the  amplitudes of the state variables of the  coupled system.
 This option may help  amplifying the smaller amplitude of the attractors noticeable or, on the contrary, limit an almost unbounded growth of an attractor by attenuating it. We make use of this
possibility to change the  amplitude of the emerging attractors in the state of EM. To demonstrate these novel features of EM we use two coupled R\"ossler  oscillators as a paradigmatic  model.
 \par So far, EM has only been demonstrated in coupled
identical oscillators which are difficult to realize in experiment as shown in \cite{Patel}. To provide a theoretical understanding of an experimental setup one has to consider slightly different oscillators and to show that EM persists  under parameter mismatch.  To this end, we test the robustness of the phenomenon of extreme multistability with respect to mismatch of parameters.\\
The paper is  organized as follows: In  Sec. \ref{scheme} we describe the
 coupling scheme to achieve EM and extend it to partial AS and MS states  as well as scaling the  amplitude of attractors. We illustrate our method of creating EM with partial AS  using numerical simulation of the R\"ossler oscillator in Sec. \ref{example}.  Section \ref{mismatch} deals with the effect of the parameter mismatch and finally we discuss our results in Sec. \ref{discussion} . \\

\section{Coupling design and Choice of Controllers}
\label{scheme}
We recall briefly the main ideas of the systematic design of coupling to achieve EM (for details cf. \cite{Hens}) and extend it to partial AS and MS.
The basic condition for the emergence of EM is: {\it at least  one pair of variables maintains a constant distance that emerges as a conserved quantity which is sensitive  to changes in the  initial state.} To obtain  partial AS we require all other variables to be antisynchronized, while for  MS some pairs of variables should be in AS and the remaining ones should be in CS.
For this  purpose of generalization, we now introduce a scalar matrix $\boldsymbol\alpha$ \cite{Padmanaban} in the definition of the error vectors of the coupled system.  We start with two identical oscillators as described by
$ \bf{\dot{x}}= \bf{F}(\bf{x}); \bf{x}= $$ [x_i; i=1,2...,n] $ and $ \bf{\dot{y}}=\bf{F}(\bf{y}), \bf{y}= $$[y_i; i=1,2...,n]$,
where $ \bf{F}: R^n \rightarrow R^n $. Assume the  synchronized dynamics  taking place on a manifold
defined by ${\bf x}={\boldsymbol\alpha\bf y}$ and the deviation from the synchronization manifold is described by the
error $ \bf{e}= \bf{x}-\boldsymbol\alpha\bf{y} $. The constant matrix $\boldsymbol\alpha$ ($n\times n$) implements scaling up or down the comparative  amplitude of the pairs of the variables of the coupled system \cite {Padmanaban}  and thereby takes care of increasing or decreasing the  amplitude of the attractors.  Additionally, it takes care of the type of synchronization.
If the matrix $\boldsymbol\alpha$  is the identity matrix then all the variables will be completely synchronized.
However, all the elements $\alpha_{ij}$ ($i,j=1,....,n$) of the $\boldsymbol\alpha$-matrix may not be positive natural numbers. They may take any real value ($\alpha_{ij}\in \bf{R^1}$). For simplicity we focus here only on the matrices $\boldsymbol\alpha$ which are diagonal. If the diagonal elements of the $\boldsymbol\alpha$-matrix are all negative but the values are -1, then the variables are in an AS state. The  MS state can be created if some of the diagonal elements of the $\boldsymbol\alpha$-matrix are chosen to be +1 and some others are   -1. Choosing  the elements  larger or smaller than unity, allows  amplification or attenuation of the amplitude of the attractors of one system compared to another when the systems are coupled.
The final form of  the error dynamics is described as:
 $\dot{\bf{e}} = \bf{G}(\bf{x},\bf{y},\boldsymbol\alpha)=\bf{F}(\bf{x})-\boldsymbol\alpha\bf{F}(\bf{y})$, $ \bf{G}: R^n\rightarrow R^n $.
To fulfill the above mentioned conditions, the error dynamics has to obey a specific targeted form which we denote by $\tilde{\bf{G}}(\bf{x}, \bf{y},\boldsymbol\alpha)$.
This desired error dynamics $\tilde{\bf{G}}(\bf{x}, \bf{y},\boldsymbol\alpha)$ is realized by designing a set of controllers ${\bf u_1}(\bf{x},\bf{y})$ and ${\bf u_2}(\bf{x},\bf{y})$
for coupling the two dynamical systems in such a way that it originates EM in the coupled system. Now the desired error dynamics  $\dot{\bf{e}} =  \bf{\dot{x}} -\boldsymbol\alpha\bf{\dot{y}}=\tilde{\bf{G}}(\bf{x},\bf{y},\boldsymbol\alpha) $ implies that $\tilde{\bf{G}}(\bf{x},\bf{y},\boldsymbol\alpha)-F(\bf{x})+ \boldsymbol\alpha F(\bf{y}) = \bf{u_1(\bf{x},\bf{y})}-\boldsymbol\alpha\bf{u_2(\bf{x},\bf{y}).} $
According to a particular choice of the matrix $\boldsymbol\alpha$, one can choose the controllers $\bf{u_1}$ and $\bf{u_2}$ in such a way that CS is obtained in {\it p}-pairs of variables  and AS in the  ({\it n-1-p})-pairs of variables and, the remaining one pair maintains a constant distance. In this  way any possible state, either a partial MS  or a  partial CS or AS can be realized.  This generalizes the concept introduced
in \cite{Hens} to different kinds of synchronization and  additionally, introduces an option to amplify or attenuate the  amplitude of the attractors.
\par As already mentioned in \cite{Hens}, the choice of controllers is not unique, but very flexible. Here we quantify the flexibility of the choice of controllers. To derive the minimum number of different controller choices when coupling two n-dimensional oscillators, we start with  3-dimensional systems. First of all we mention that it is necessary to have a mutual coupling
 to achieve EM. For this reason we cannot set all the controllers of one oscillator to zero, i.e. $u_{1i}=0$ or $u_{2i}=0$ $\forall i$ is  forbidden. This leaves us with the possible choices (1) $u_{12}=0$; $u_{21}=0$ and $u_{23}=0$, (2) $u_{11}=0$; $u_{22}=0$ and $u_{23}=0$ and (3) $u_{13}=0$; $u_{21}=0$ and $u_{22}=0$, where one controller of the first oscillator and two controllers of the second oscillator are set to zero. Alternatively we can also nullify two controllers of the first oscillator and one controller of the second oscillator which yields another 3 choices. Thus we can construct $3+3=6$ different  types of controllers for  one particular  choice of error dynamics in 3D systems.
 Next we consider a 4-dimensional system for which we obtain $ {{4}\choose{1}}+{{4}\choose{2}}+{{4}\choose{3}}=14$  choices of controllers  from a simple counting of the possible choices for one specific error dynamics. In a similar manner,  we determine the total number of choices of controllers for  couplig  two n-dimensional  systems are $ {{n}\choose{1}}+ {{n}\choose{2}} +{{n}\choose{3}}+....+{{n}\choose{n-1}}=2^n-2$.
This follows from the relation  $(1+x)^n=1+{{n}\choose{1}}x+{{n}\choose{2}}x^2+{{n}\choose{3}}x^3+....+x^n$ with setting   $x=1$.  We emphasize that this minimal number of controllers is available for one particular choice of the error dynamics.  
 
 However we can also choose  $\boldsymbol u_1=\wp[\tilde{\bf{G}}(\bf{x},\bf{y},\boldsymbol\alpha)-F(\bf{x})+\boldsymbol\alpha F(\bf{y})]$ and $\boldsymbol u_2=-(1-\wp)[\tilde{\bf{G}}(\bf{x},\bf{y},\boldsymbol\alpha)-F(\bf{x})+\boldsymbol\alpha F(\bf{y})]$ with $\wp$ being any rational number between 0 and 1. As a consequence, in principle, infinitely many choices of controllers are available for coupling two systems to realize EM.  This large coupling options opens up   possibilities of a physical realization of the EM.

\section{Extreme Multistability: Numerical Examples}
\label{example}
To illustrate the emergence of EM in a partial AS state,   we consider  two identical R\"ossler systems coupled through  bi-directional controllers  given by
\begin{figure*}
\vspace{2pt}
\begin{center}
\centering
\subfigure(a)\label{rossler_attra_gen}\includegraphics[width=0.45\textwidth]{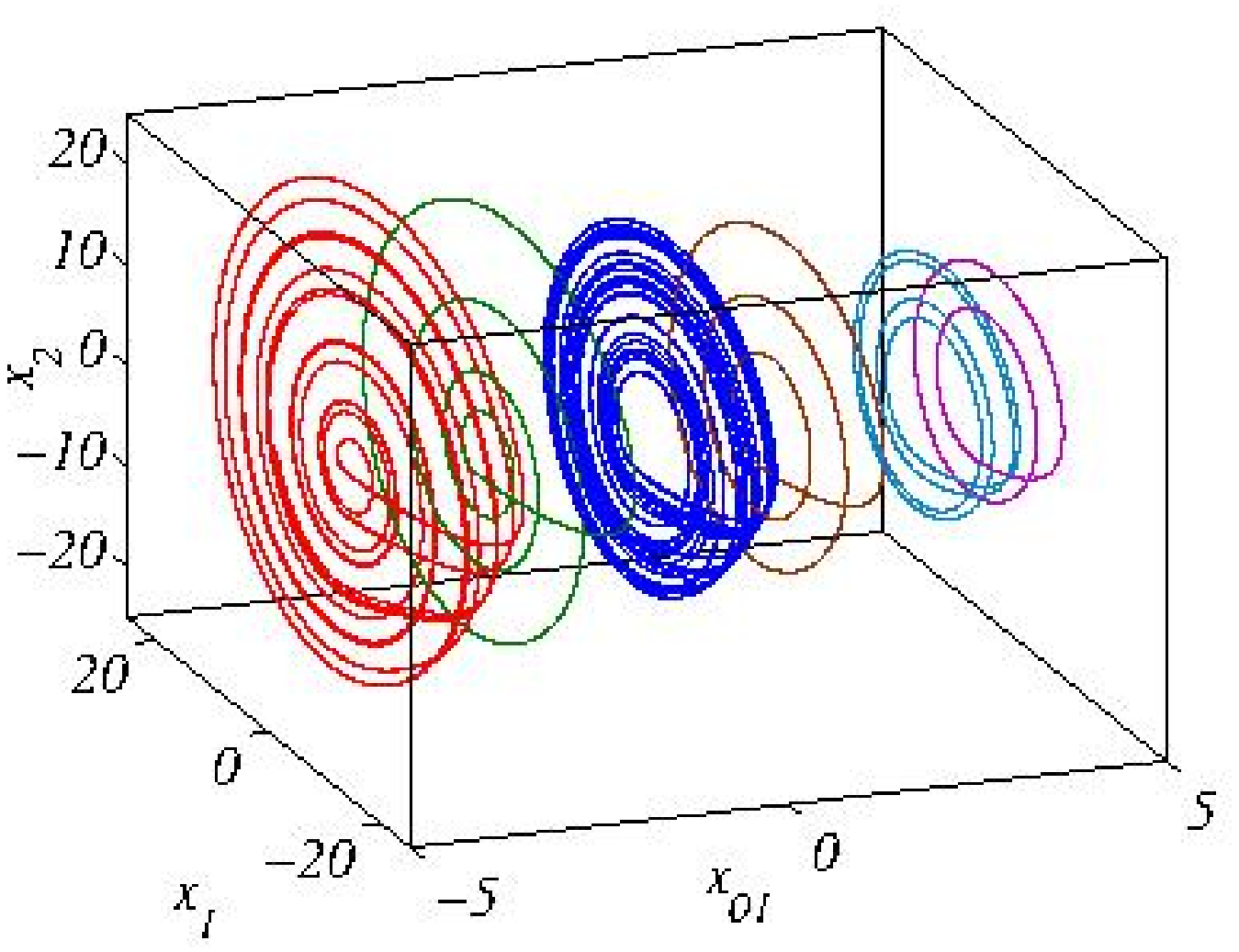}
\subfigure(b)\label{rossler_attra_gen}\includegraphics[width=0.45\textwidth]{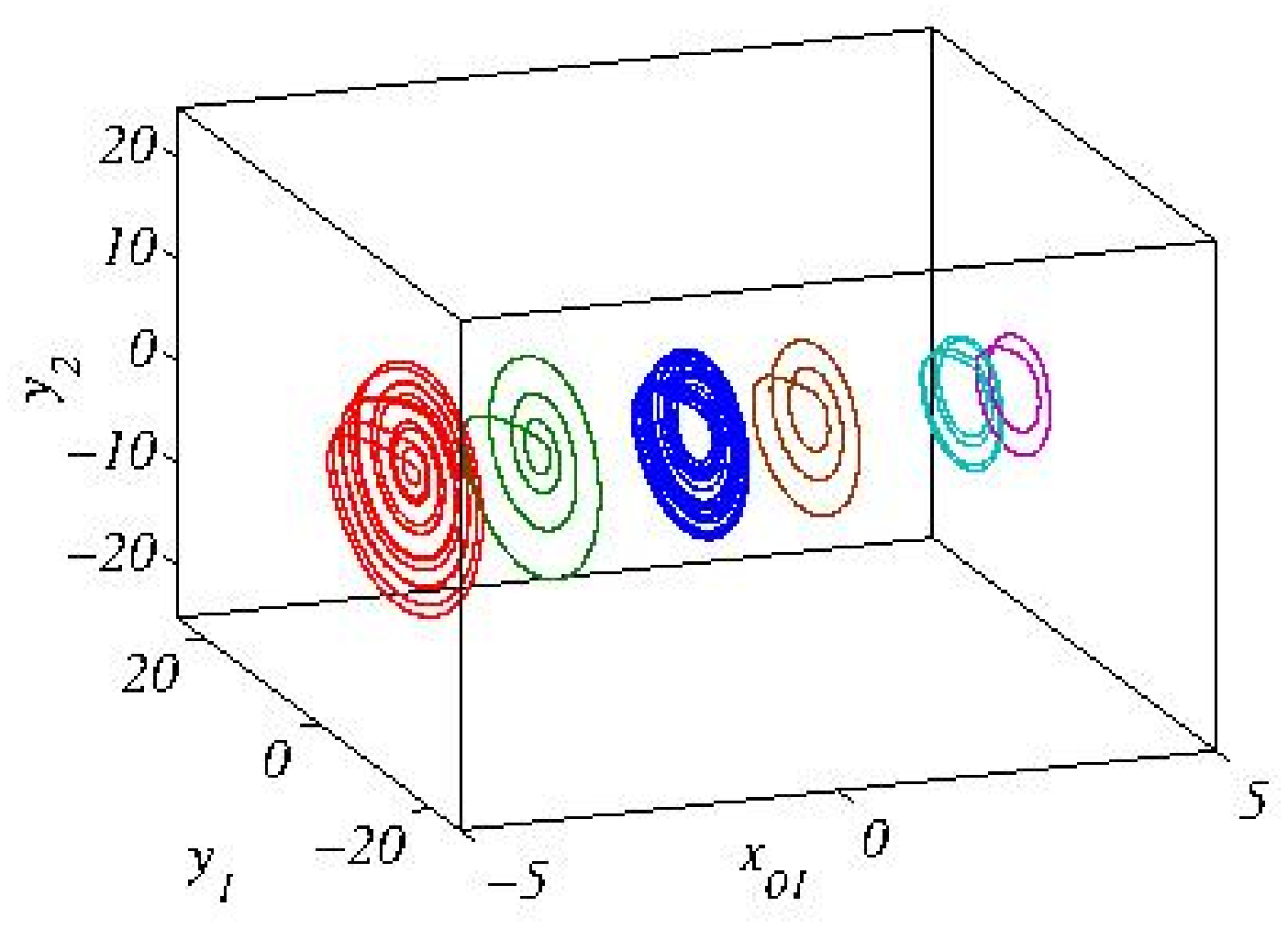}
\end{center}
  \caption{Multiattractor dynamics in coupled R\"ossler systems: 2D projection of different trajectories converging to different attractors for
different initial states, $x_{01}$. (a) $x_1$ vs $x_2$ and (b) $y_1$ vs $y_2$.  $\alpha_{11}=-2.0$ ,$\alpha_{22}=-2.0$ and $\alpha_{33}=-2.0$.
 For comparison, two figures are drawn in the same scale. It confirms that one set of attractors is an amplified version of the other set and  they are in antiphase. Other initial conditions are  fixed at $x_{02}=0.0; x_{03}=0.2; y_{01}=-0.1; y_{02}=-0.1; y_{03}=0.0;$
$x_{01}$ has only been changed from  -4 to 4.} \label{rossler_attra_gen}
\end{figure*}

\begin{subequations}
\label{Rossler_as}
\begin{eqnarray}
\dot{x}_1&=&-x_2-x_3+u_{11}, \\
 \dot{x}_2&=&x_1+a x_2+u_{12},\\ 
\dot{x}_3&=&b+x_3(x_1-c)+u_{13}, \\ 
 \dot{y}_1&=&-y_2-y_3+u_{21},  \\
\dot{y}_2&=&y_1+a y_2+u_{22}, \\
\dot{y}_3&=&b+y_3(y_1-c)+u_{23} .
\end{eqnarray}
\end{subequations}
$u_{ij}$  (i=1, 2 and j=1, 2, 3)  are controllers which are to be designed to define the coupling.
The parameters of the uncoupled R\"ossler systems are set  to such values that a chaotic  attractor ($a=0.2;b=0.2;c=5.7$) exists; no multistability occurs.
 We can now derive the coupling function by an appropriate choice of the controllers to realize EM with a partial AS or MS along with mutual amplification or
attenuation of the coexisting attractors. We have already introduced  the scaling matrix in the definition of the errors in the previous section.
 The errors have been defined by $ e_i = x_i-\alpha_{ii}y_i$ where $\boldsymbol\alpha$ is a diagonal matrix (for simplicity)
and its elements $\alpha_{ii} (i=1, 2, 3)$ are real numbers. 
 Once a desired synchronization regime is targeted by a proper selection of  $\alpha_{ii}$,
we concentrate on the choice of controllers for realizing  EM. We design the error dynamics in
such a way that the coupled system is partially synchronized in any of the desired synchronization regimes (CS, AS or MS)
in which,  at least,  one pair of variables evolves into a conserved quantity. For the present example we choose the desired error dynamics as
 \begin{eqnarray}
\label{error_dynamics}
  {\dot e}_1 &=& -e_3 + y_1 e_2, \nonumber \\
  {\dot e}_2 &=& x_2 e_3, \label{errors} \\
  {\dot e}_3 &=& -c e_3-x_2 e_2. \nonumber
\end{eqnarray} For a choice of  Lyapunov function $ V=e_2^2+e_3^2 $,  $e_2$ and $e_3$ tend to zero in the long run while $e_1$ emerges as a constant $C$ depending on the initial conditions. Note that this choice of  the error dynamics and the Lyapunov function is not unique as disucssed above.
We can always make other choices provided they satisfy the basic conditions of partial synchrony and  introduces an emergent conserved quantity depending upon the initial state.
Using the difference equations of \ref{Rossler_as}(a)-\ref{Rossler_as}(d), \ref{Rossler_as}(b)-\ref{Rossler_as}(e), \ref{Rossler_as}(c)-\ref{Rossler_as}(f), and  the error dynamics  (\ref{errors}) we derive the controllers as follows
 \begin{eqnarray}
\label{controllers}
u_{11} &=& y_1 e_2 + x_2-\alpha_1 y_2 +x_3-\alpha_1 y_3 -e_3, \nonumber\\
u_{12} &=& 0, \nonumber\\
u_{13} &=& -x_2e_2-b+\alpha_3 b-x_1 x_3 +\alpha_3 y_1 y_3, \label{controller}\\
u_{21} &=& 0, \nonumber\\
u_{22} &=& \frac{x_2 e_3 -x_1 +\alpha_2 y_1 -a e_2}{-\alpha_2}, \nonumber\\
u_{23} &=& 0. \nonumber
\end{eqnarray}

Using these choices of the controllers, we obtain the coupled   R\"ossler oscillators  that exhibit  EM,
  \begin{eqnarray}
    {\dot x}_1 &=&-x_2-x_3 + y_1 (x_2-\alpha_2 y_2) + x_2-\alpha_1 y_2 \nonumber\\
& & +x_3-\alpha_1 y_3 - x_3- \alpha_3 y_3 , \nonumber\\
   {\dot x}_2 &=& x_1+ax_2,\nonumber\\
   {\dot x}_3 &=& b+x_3(x_1-c)-x_2(x_2-\alpha_2 y_2)-b+\alpha_3 b-x_1 x_3\nonumber\\
& & +\alpha_3 y_1 y_3,\label{rossler}\\
    {\dot y}_1 &=& -y_2-y_3 , \nonumber\\
    {\dot y}_2 &=& y_1+ay_2 \nonumber\\
& &+ \frac{x_2(x_3-\alpha_3 y_3) -x_1 +\alpha_2 y_1 -a (x_2-\alpha_2 y_2)}{-\alpha_2}, \nonumber\\
    {\dot y}_3 &=& b+x_3(x_1-c).\nonumber
\end{eqnarray}

\begin{figure*}
 
\vspace{5pt}
 \begin{center}
\centering
\subfigure(a)\label{rossler_bifur_gen}\includegraphics[width=0.45\textwidth]{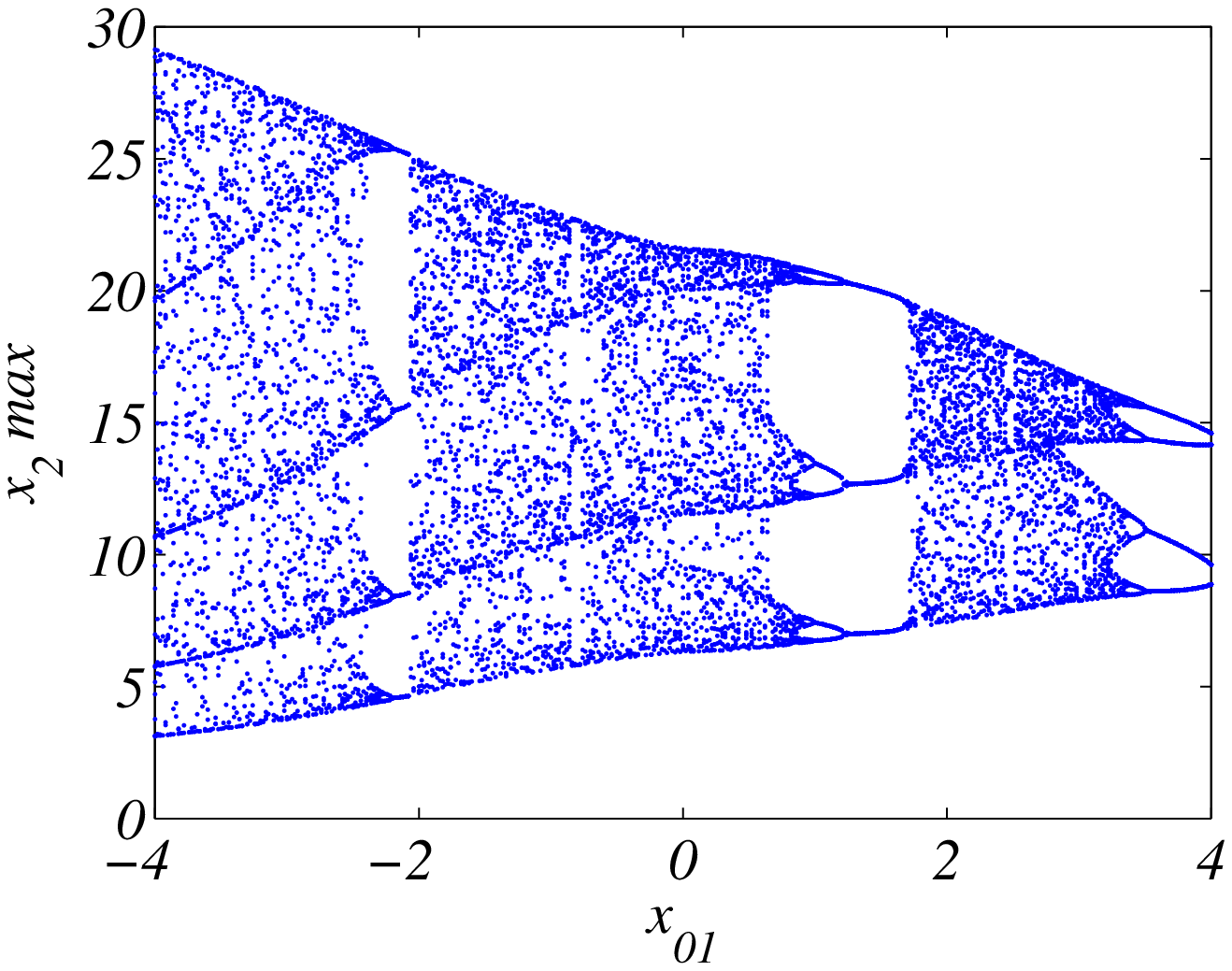}
 \subfigure(b)\label{rossler_constant}\includegraphics[width=0.45\textwidth]{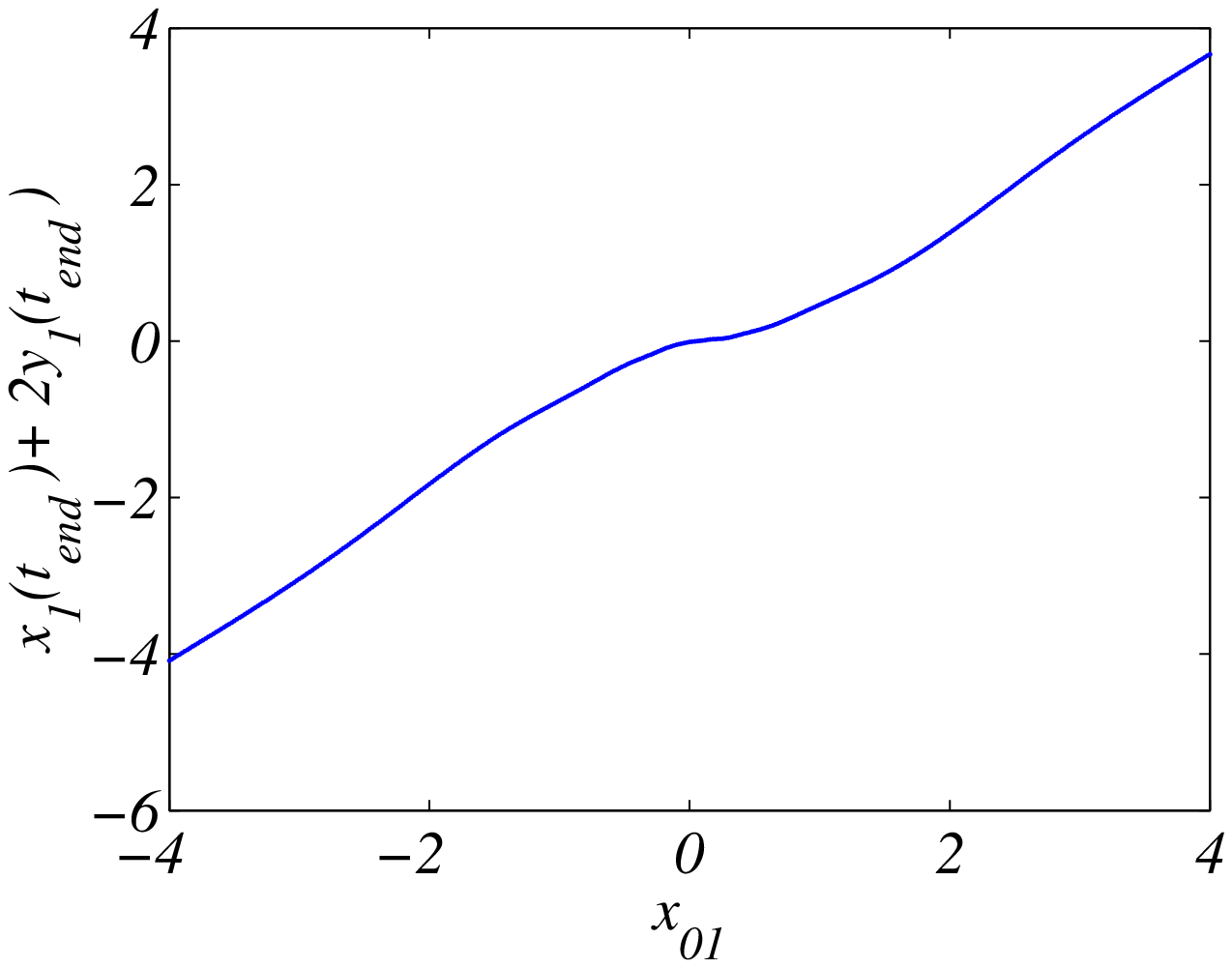}
\end{center}
  \caption{ (a) Maxima values of $x_2$ as a function of initial condition $x_{01}$. The rest initial states are fixed as said before. (b) An emerging constant as a function of same initial condition $x_{01}$.} \label{rossler_bifur_gen_const}
\end{figure*}
\

 The errors $e_2$ and $e_3$ go to zero in the long run since $\dot{V}=-ce^2_{3}$, $e_1$ maintains a constant distance. 
For a choice of $\alpha_{11}=-2.0$, $\alpha_{22}=-2.0$ and $\alpha_{33}=-2.0$, all the states are in antiphase and one set of attractors is expected to be larger by twice the other set of attractors. The conserved quantity still appears as a sum of two variables $e_1=x_1+2y_1=e_1^*$ ($e_1^*$ is the conserved  quantity).

Figures \ref{rossler_attra_gen}(a)-(b) show the attractors ($x_1$ vs $ x_2$ and $y_1$ vs $y_2$ plots) of the coupled system
with  variation of one initial state $x_{01}$. They clearly reveal the coexistence of a large number of attractors which are in AS state. The  mutual amplification of the attractors (one is  twice as larger as the other) is also clearly seen by comparison of the Figs. \ref{rossler_attra_gen}(a) and \ref{rossler_attra_gen}(b).

The infinite number of coexisting attractors is in fact  confirmed by the bifurcation-like diagram in  Fig. \ref{rossler_bifur_gen_const}(a) by plotting
the maxima or peak values of  $x_{2}$ for changing initial condition $x_{01}$. Many dynamical states (limit cycle or chaotic) appear for each initial state  $x_{01}$.  The emerging constant ($x_1$($t_{end}$)+2$y_1$($t_{end}$)) is also plotted versus $x_{01}$ in  Fig. \ref{rossler_bifur_gen_const}(b). Here  $t_{end}$ is the final integration time which is chosen to be $t_{end}=500000$.  This confirms that the emergent constant follows a nonlinear relation with the initial state and thus it is challenging to predict the emergent attractor for a chosen set of initial conditions.  We do not elaborate the MS state here with an example, since it is clear that, by using a different set of $\alpha_{ii}$, one may ensure MS which does not change the main algorithm for realization of the EM.

\section{Robustness of EM to parameter mismatch}
\label{mismatch}
 Finally we check the effect of parameter mismatch on  EM. Since no two oscillators, in the real world, are exactly identical, there exists a mismatch in their parameters. 
This poses a problem for the experimental realization of EM by coupling two oscillators with a mismatch which may finally lead to the disappearance of EM. Recently, EM has been evidenced in a laboratory experiment
\cite{Patel} using two almost identical electronic analogs of R\"ossler oscillators employing a simple error dynamics \cite{Hens}. It was found that, as expected, the dynamics  drifts continuously from one attractor to another  qualitatively different attractor. This "drifting" of attractors appears due to an instability created by the parameter mismatch in the coupled oscillators. 
It was observed that if the coupled oscillators were fabricated as closely identical as possible, each attractor remains stable for a set of initial conditions and for a reasonably long time before drifting to another attractor. The individual attractor remains stable for a longer time and the drifting is slower when the mismatch is much smaller. On the contrary the drifting becomes faster for increasingly larger mismatch. This clearly indicates the existence of EM in a coupled system. Thus, given two closely identical systems, one can really observe various stable attractors for a change of initial conditions.  On the other hand, this parameter mismatch could be taken care of  from the beginning by giving a special  attention to the design of coupling. By this way, one can obtain stable attractors avoiding the  effect of drifting.
For a demonstration, we consider two R\"ossler oscillators with non-identical parameters. The first oscillator has parameters $a, b, c$.
\begin{subequations}
\label{ross_match_1}
\begin{eqnarray}
{\dot x}_1&=&-x_2-x_3,\\
{\dot x}_2&=&x_1+ax_2, \\
{\dot x}_3&=&b+x_3(x_1-c).
\end{eqnarray}
\end{subequations}
The second  oscillator has been detuned by $\Delta$ mismatch,
\begin{subequations}
\label{ross_match_2}
\begin{eqnarray}
{\dot y}_1&=&-y_2-y_3,      \\
 {\dot y}_2&=&y_1+(a+\Delta_a)y_2, \\
 {\dot y}_3&=&b+\Delta_b+y_3(y_1-c-\Delta_c).
\end{eqnarray}
\end{subequations}
We again fix the  parameters  in the chaotic regime ($a=0.2,b=0.2, c=5.7$) for both  oscillators.

\begin{figure}

\vspace{2pt}
\begin{center}

\centering
\subfigure(a){\label{rossler_bifur_mismatch}\includegraphics[width=0.4\textwidth]{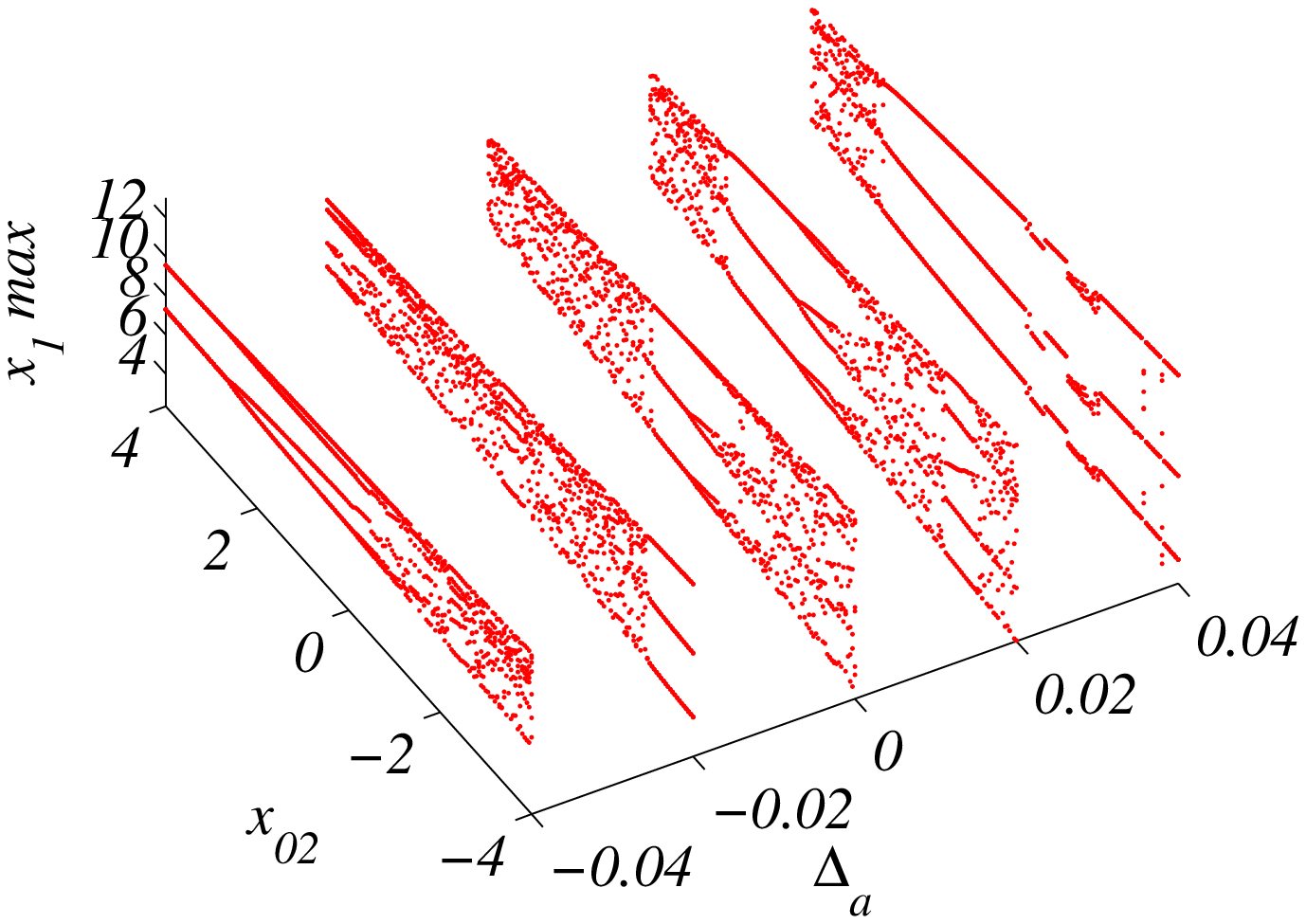}}
\subfigure(b){\label{rossler_consta_mismatch}\includegraphics[width=0.4\textwidth]{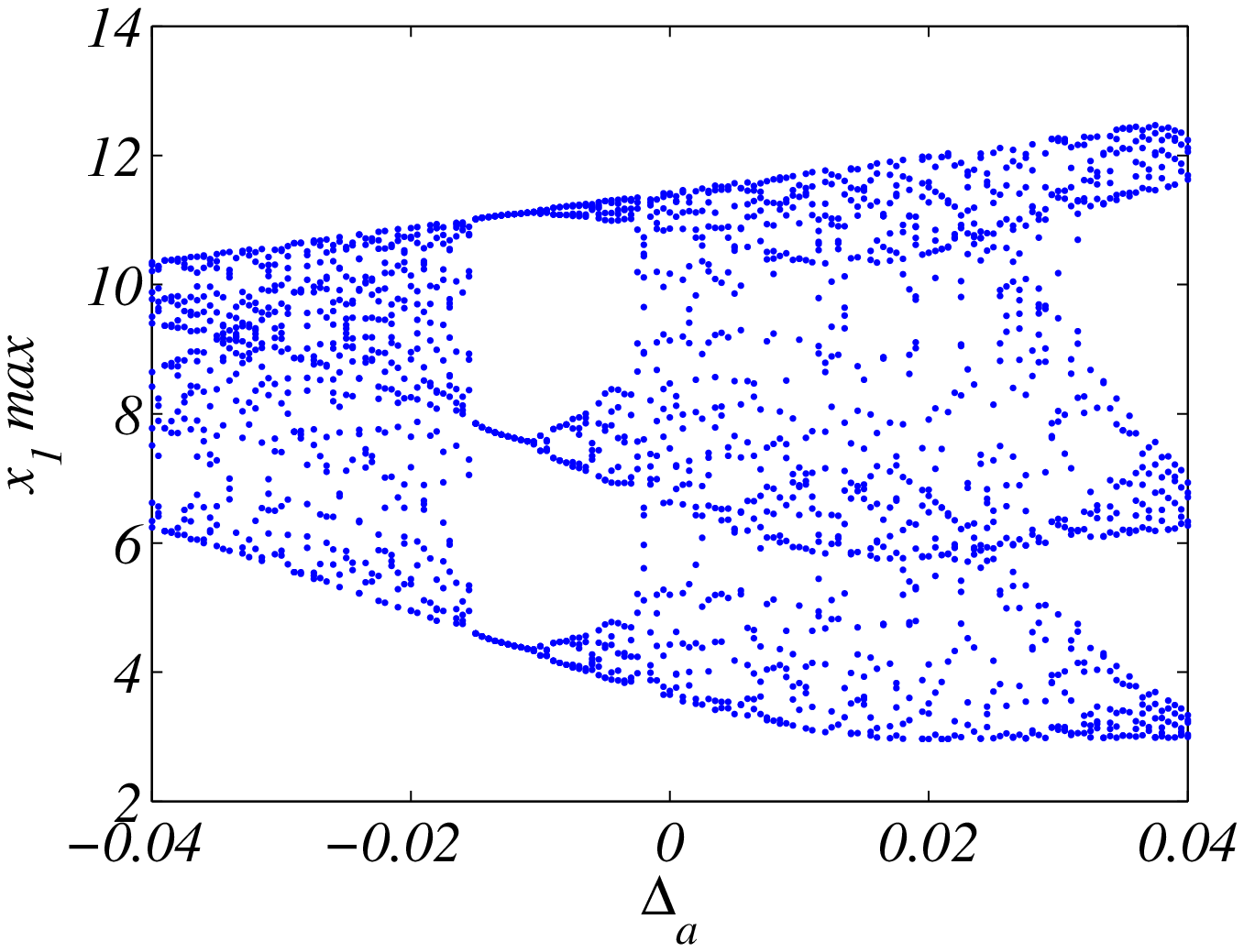}}
\hspace{-0.05in}

\end{center}
 \caption{(a) Maxima of $x_1$ as a function of initial condition $x_{02}$ and   mismatch $\Delta_a$,
$\Delta_a =-0.04;-0.02; 0 ; 0.02;0.04$. Other initial conditions are set to
$x_{01}=0.0;x_{03}=-0.3;y_{01}=3.1;y_{02}=-2.0;y_{03}=-0.1$. (b) Maxima of $x_1$ as a function of mismatch $\Delta_a$, when $x_{02}$ is fixed at 2.0.} 
\label{rossler_bifur_mis_const}
\vspace{5pt}
\end{figure}
To gain  more insight into the parameter mismatch we analyze the general setup elaborated in section II. Let an $n$-dimensional system  be  governed by the equations $ {\mathit{\bf{\dot{x}}}}= \bf{F}(\bf{x},\boldsymbol\mu) $;  $ \bf{F}: R^n \rightarrow R^n $; where  $\boldsymbol\mu$ depicts all the parameters of the system. Another uncoupled system whose parameters are different from the first one is described by
$ \bf{\dot{y}}= \bf{F}(\bf{y},\boldsymbol\mu') $ and all the parameters of this system are denoted  by $\boldsymbol\mu'$ with $\boldsymbol\mu'=\boldsymbol\mu+\boldsymbol\Delta_{\mu}$. When coupling these two systems,
 we define  the errors between the variables as $\bf{e} =\bf{x}-\boldsymbol\alpha\bf{y}$ where the description of ${\boldsymbol\alpha}$ is the same as before. However, the  error dynamics is governed by  $\dot{\bf{e}} = \bf{F}(\bf{x,\boldsymbol\mu})-{\boldsymbol\alpha{F}}(\bf{y,\boldsymbol\mu'}) +u_{1}-{\boldsymbol \alpha u_{2}}$.\\
 Our  target is that the detuning of parameters cannot destroy the synchronization manifold as well as the complex behaviour of the coupled system, namely, the EM.
Suppose we can construct an error dynamics in the following form
$\dot{\bf{e}}=(\boldsymbol\mu -\boldsymbol\mu'){\bf{H}}(\bf{x},\bf{y})$. By designing the controllers in such a way that they depend on the mismatch, we are able to ensure EM for rather large intervals of mismatch. Moreover the deviation from the parameter of the first oscillator can be either positive or negative.

The nonidentical R\"ossler oscillators  have already been described by the equations \ref{ross_match_1}(a)-(c) and \ref{ross_match_2}(a)-(c).
For simplicity  we are dealing with an error dynamics for which  we set all the $\alpha_{ii}$ to 1, i.e.,   $e_i=x_i-y_i, i=1,2,3$.
One choice of the error dynamics is,
${\dot e}_1= x_2 e_3$, 
${\dot e}_2= \Delta_a e_1$  and 
${\dot e}_3=-x_2e_1-c e_3$.  
One parameter ('a') exists in the equation of  $\dot{x}_2$, the other two ('b' and 'c') are connected with $\dot{x}_3$.
For simplicity we have introduced the mismatch only in parameter $a$. Therefore $\Delta_b=\Delta_c=0.0$.

We have taken a Lyapunov function $V={\frac{1}{2}}(e_1^2+e_3^2)>0$ for
which $\dot{V}=e_1 \dot{e_1}+e_3 \dot{e_3}=-c e_3^2$ which ensures global stability for $c>0$.
  Following the above mentioned condition, $e_2$ becomes constant $C$ for any real values of  $\Delta_a$.
 The differences of the controllers can be obtained from the differences of  \ref{ross_match_1}(a)-\ref{ross_match_2}(a),
\ref{ross_match_1}(b)-\ref{ross_match_2}(b), \ref{ross_match_1}(c)-\ref{ross_match_2}(c) following the error dynamics,
   \begin{eqnarray}
      u_{11}-u_{21}&=& -y_2-y_3+x_2+x_3+x_2(x_3-y_3),\nonumber \\
    u_{12}-u_{22}&=&-e_1 -a e_2+\Delta_a e_1+\Delta_a y_2,\\
    u_{13}-u_{23}&=&-x_3(x_1-c)+y_3(y_1-c)\nonumber\\
&& -c(x_3-y_3)-x_2(x_1-y_1).
  \end{eqnarray}
We set $u_{11}$, $u_{22}$ and $u_{23}$ to zero.

Then the coupled  equations become
\begin{subequations}
\label{coupled_mismatched rossler}
\begin{eqnarray}
{\dot{x}}_1&=&-x_2-x_3, \\
{\dot{x}}_2&=&x_1+a x_2+((\Delta_a-1)(x_1-y_1)-a(x_2-y_2)+\Delta_a y_2),\\
{\dot{x}}_3&=&b+(x_1-c)x_3+(-x_3(x_1-c)+y_3(y_1-c)\nonumber \\
&&-c(x_3-y_3)-x_2(x_1-y_1)), \\
{\dot{y}}_1&=&-y_2-y_3-(x_2(x_3-y_3)+x_2-y_2+x_3-y_3),\\ 
{\dot{y}}_2&=&y_1+(a +\Delta_a)y_2, \\
{\dot{y}}_3&=&b+y_3(y_1-c).
\end{eqnarray}
\end{subequations}
 We have plotted the maxima of $x_2$ in Fig. \ref{rossler_bifur_mis_const}(a) showing the bifurcation like sequence obtained by varying the initial value $x_{02}$ from -4 to +4 for different values of the mismatch  ($\Delta_a=-0.04;-0.02;0;0.02,0.04$).  Each sequence for a particular parameter mismatch $\Delta_a$ in Fig. \ref{rossler_bifur_mis_const}(a) illustrates the fact that there exists a unique attractor (chaotic or limit cycle) for each value of $x_{02}$. It is clearly seen that the EM is not destroyed by the parameter mismatch, positive or negative. However, the particular attractor found for a certain initial condition $x_{02}$ depends on the value of the mismatch as shown in Fig. \ref{rossler_bifur_mis_const}(b). We  mention that robustness with respect to parameter mismatch is established with this example in which  a partial CS is achieved by the choice of $\alpha_{ii}=1$. In general, the basic framework of the coupling configuration would, of course, also allow for other choice of $\alpha$ such as -2  to establish a partial AS state with an amplification of the attractor. This choice does not disturb the emergence of EM and its robustness as well.

\section{Discussion}
\label{discussion}
We investigated the phenomenon of EM that reveals coexistence of infinitely many attractors in two coupled oscillators. EM is manifested in  coupled $n$-dimensional systems with two basic properties: (1)  partial synchronization (CS or AS or MS) of $n-1$ components (state variables) while the $n-th$ components keep a constant distance depending on the initial conditions, (2)  emergence of a conserved quantity that characterizes the synchronization manifold in which the attractor lives. In the long-term limit the state space appears as foliated into infinitely many synchronization manifolds, each of them containing at least one attractor for each emergent constant determined by the set of  initial conditions.  As explained in detail in the introduction, we emphasize here again, that all attractors studied in this paper are in fact relative attractors, since they are attractors within the synchronization manifold in which they exist, but marginally stable in the direction perpendicular to the synchronization manifold. Our investigation is based on the derivation of controllers that allow an appropriate design of coupling leading to the state of EM. In this paper, this controller design, which has been introduced earlier in \cite{Hens}, has been extended to achieve partial AS or MS in which some, say $m$, of the components are completely synchronized while $n-m-1$ components are antisynchronized. This generalization of the coupling scheme which is based on Lyapunov function stability is also employed to amplify or attenuate the   amplitude of the attractors of the coupled system. This property could be useful in an experimental set up leading to either very large or very small amplitude attractors.
 We have shown that the choice of controllers is wide and have given an analytical estimate of the possible number of different coupling schemes which could realize the desired extreme multistability.  For each choice of the Lyapunov function and  each possible choice of nullifying the differences between the controllers of the two coupled systems, we can get at least $2^n-2$ different designs of the controllers. We have shown that  infinitely many  options  of coupling functions are available for realizing EM in two coupled systems. This offers a great flexibility, particularly, in view of an experimental design, since experimental setups might have some restrictions on a possible coupling design. Offering different coupling options always enhances the possibility of an experimental realization of the EM.

Additionally, again with respect to experimental design, we addressed the robustness of EM  to parameter mismatch since, in reality, oscillators will never be identical as usually assumed for theoretical investigations. It turns out that, with a proper design of the controllers, a larger mismatch is allowed without disturbing the basic conditions to sustain EM  in coupled systems.

The controller design has thus been shown to be very effective in realizing EM and very flexible. When EM emerges in two coupled systems, a conserved quantity appears in the long-term limit. This conserved quantity depends on the initial condition of the two coupled systems and determines the synchronization manifold in which a particular attractor lives. However, it is still an open question, how this controller design leads to simple or complex dependence of the conserved quantity on the initial conditions. To clarify this question could be subject of further research.

\section{Acknowledgements}
The authors acknowledge Ravindra E. Amritkar, Abhijit Sen and Ken Showalter for interesting discussions.  Particularly we would like to thank an unknown reviewer for his substantial comments and James A. Yorke for illuminating insights. C.R.H  and S.K.D. acknowledge support by the CSIR Emeritus scientist scheme, India. U.F. would like to thank Rajarshi Roy and his group for their hospitality and acknowledges support from Burgers Program for Fluid Mechanics of the University of Maryland at College Park.


\begin{thebibliography}{28}
\bibitem{explain} The concept of relative attractors has been introduced in the hydrodynamical context \cite{Tel,Skufca} and differs slightly from our situation because there the attractor is unstable in the perpendicular direction.
\end{thebibliography}
\end{document}